\documentclass[english]{article}
\usepackage[T1]{fontenc}
\usepackage[latin9]{inputenc}
\usepackage{babel}
\usepackage{amssymb}
\usepackage{esint}
\usepackage[unicode=true,pdfusetitle,
 bookmarks=true,bookmarksnumbered=true,bookmarksopen=false,
 breaklinks=false,pdfborder={0 0 1},backref=section,colorlinks=false]
 {hyperref}
\begin{document}

\title{An algebraic approach to systems with dynamical constraints }

\author{Jerzy Han\'{c}kowiak}

\author{(former lecturer and research worker of Wroclaw and Zielona Gora
Universities)}

\author{Poland}

\author{EU}

\author{e-mail: hanckowiak@wp.pl}

\date{October 2013}
\maketitle
\begin{abstract}
Constraints imposed directly on accelerations of the system leading
to the relation of constants of motion with appropriate local projectors
occurring in the derived equations are considered. In this way a generalization
of the Noether's theorem and a relation of local quantities to global
are highlighted. A phenomenon of nonphysical degrees of freedom is
also discussed. 


\tableofcontents{}
\end{abstract}

\section{Introduction}

We consider equations describing discrete or continuous systems with
constraints. If no constraints are present, we will assume that the
\textit{unconstrained system} is described by the 'field' equation:

\begin{equation}
L[\tilde{x};\varphi]+\lambda N[\tilde{x};\varphi]+G(\tilde{x})=0\label{eq:1}
\end{equation}
 with the main linear functional $L$ depending on the unknown 'field'
(function) $\varphi(\tilde{x})$, which necessarily includes differential
operations, for example 

\[
L[\tilde{x};\varphi]=\left(\square+m^{2}\right)\varphi(\tilde{x})
\]
the $N$ a functional,\textbf{ usually} nonlinear (although may also
contain additional linear terms), depending on the field $\varphi$,
for example

\[
N[\tilde{x};\varphi]=\varphi^{3}(\tilde{x})
\]
and the given function $G$ usually describing external forces acting
on the system. Here and further square brackets mean that a given
quantity, except that it is a function, it is also a functional. For
discrete systems, such as N material points, $\varphi=(q_{1},...,q_{3N})$
can be a 3N dimensional vector and $\tilde{x}=(t,i)$ besides the
time $t$ describes the component indexes $i=1,...,3N$. In this case,
we can choose $L[t,i;\varphi]=\ddot{q_{i}}(t)$. In general, the 'vector'
$\tilde{x}\in\tilde{M}$ has time-'space' components describing 'points',
components characterizing the field $\varphi$ as its tensor type,
and the time $t$. Usually, we will distinguish the time and 'space'
components by writing, e.g., $\tilde{x}=(t,\bar{x})$. We will assume
that all components of the vector $\tilde{x}$ are discrete variables.
In other words,$\tilde{M}$ is a set defined by a specific properties
of the considered system, see also App.7. 

We are in good company. Even the space can be described by means of
the field $\varphi(\tilde{x})$. 

The functionals $L,N$ are also functions depending on the 'vectors'
$\tilde{x}\in\tilde{M}$. The set of functions dependent on the fixed
$\varphi$ will be denoted by $\tilde{F}_{\varphi}$ . 

As usual, we will assume that the freedom of the theory described
by Eq.\ref{eq:1} is such as the freedom of the theory descibed by
the main linear part:

\begin{equation}
L[\tilde{x};\varphi]=0\label{eq:2}
\end{equation}
 It means that in both cases the same type of initial and boundary
conditions can be used to get the unique solution. 

We also assume, following an analogy with the classical mechanics,
see also \cite{Edwadia 1993} and App.1, that the system represented
by the field $\varphi$ - subjects to the restrictions of the following
type:

\begin{equation}
\int Q[\tilde{x},\tilde{y};\varphi]L[\tilde{y};\varphi]d\tilde{y}=f[\tilde{x};\varphi]\label{eq:3}
\end{equation}
 where $Q[\tilde{x},\tilde{y};\varphi]$ is a given projector ($Q=Q^{2}$)
acting in the linear space of functions $\tilde{F}_{\varphi}$and
$f[\tilde{x};\varphi]$ is a given  function. They both are depending
in the linear or nonlinear way on the fixed 'field' $\varphi$. The
restrictions (\ref{eq:3}) together with additional assumption imposed
on the 'reaction forces' $R[\tilde{x};\varphi]$ , such as (\ref{eq:7}),
are called here the \textbf{\textit{dynamical constraints }}(DC).
They can be ideal\textbf{ }(IC) or non-ideal constraints (NIC) as
well as holonomic or non-holonomic. 

In the paper we show how Eq.\ref{eq:1} is changing in case of NIDC,
Sec.2, and how Eq.\ref{eq:3} can be interpreted, Sec.3. In Sec.3
we also show how all formulas and equations are additionally changing
in the case of ideal constraints, see also App.3. 

In the paper the concept of virtual displacements, typical tool when
discussing systems with constraints, is replace by the algebraic concepts
such as the projection operators (projectors), see: \cite{Edwadia 1993},
\cite{Przew. 1988,Hanc 2012}. This facilitates the necessary modification
of the theory with constraints and illuminates relations of local
to global quantities of the theory.

Expecting or demanding a certain conservation law in the theory and
treating it as a constraint, the reaction forces can be considered
as a sign of a new type of interaction or a necessary modification
of already existing interaction. In a sense, it would be a contrary
proceeding to the idea of spontaneous symmetry breaking. 

What I found interesting in the present study is a connection of the
certain constants of motion with the presence of certain projection
operators in the considered equations, see Eq.\ref{eq:10}. It's like
combining constants of motion with a certain symmetry of considered
equations resulting from the Noether's theorem. Equations with projectors
as in Eq.\ref{eq:10} or Eq.\ref{eq:26-1} mean that certain changes
of functionals describing these equations do not change the whole
equations.We see in this actually a generalization of symmetries of
the equations. 

In the case of restrictions (\ref{eq:3}), Eq.\ref{eq:1} has to be
changed by 

\begin{equation}
L[\tilde{x};\varphi]+\lambda N[\tilde{x};\varphi]+G(\tilde{x})=R[\tilde{x};\varphi]\label{eq:4}
\end{equation}
 with temporarily unknown 'reaction force' $R$ (a generalization
of Lagrange's equations of the first kind). In addition, I believe
that the emphasis placed here on Lagrange's equations of the first
kind is an expression of a broader approach to the description of
the nature including space, see \cite{Nerlich 1994}, - the opposite
of any kind of reductionist approache - inspite of this that they
may be acceptable in certain cases, see all arguments behind of  Lagrange's
equations of the second kind. 

For the systems with the constraints, we can look back in such a way
that we want to modify the theory determined by measuring of local
entities as the position of the various parts, taking into account
certain global (non-local) entities, for example energy of the system.
In this and only this sense, the presented approach to classical mechanics
contains some elements of quantum mechanics. 

In the case of economics system the local and global entities are
important ingredients of many theories. In this case, Adam Smith's
the invisible hand of the market would solve all the problems of capitalism
if the constrains imposed by theory would be a result of the primary
Eq.\ref{eq:1}. Otherwise, the global rules (constraints) can be used
to modify the interaction (reaction forces) between the various actors
in the market.

As in other papers, author is using integration sign even in the case
of discrete variables.

\section{The 'reaction force' R and a principle of virtual works surrogate
(PVW(S)); non-ideal constraints}

Introducing the complementary projector $P$:

\begin{equation}
P[\tilde{x},\tilde{y};\varphi]+Q[\tilde{x},\tilde{y};\varphi]=\delta(\tilde{x},\tilde{y})\label{eq:5}
\end{equation}
where $\delta$ is Kronecker or Dirac's delta, we can express the
general solution to Eq.\ref{eq:3} as follows:

\begin{equation}
L[\tilde{x};\varphi]=f[\tilde{x};\varphi]+g[\tilde{x};\varphi]\label{eq:6}
\end{equation}
 where $f[\varphi]=Q[\varphi]f[\varphi]\in Q\tilde{F}_{\varphi}$
and $g=Pg$ is an arbitrary function from $P\tilde{F}_{\varphi}$,
see (\ref{eq:18-1}). Here and elsewhere, for example:

\[
f[\varphi]=Q[\varphi]f[\varphi]\Leftrightarrow\int d\tilde{y}Q[\tilde{x},\tilde{y};\varphi]f[\tilde{y};\varphi]
\]
 Equality (\ref{eq:6}) mean that only certain components of the acceleration
of the system are completely expressed by the field $\varphi$. 

Assuming that the 'reaction forces' are such that 

\begin{equation}
\int d\bar{y}P[\tilde{x},\tilde{y};\varphi]R[\tilde{y};\varphi]=0\label{eq:7}
\end{equation}
 we get from Eq.\ref{eq:4} that

\begin{equation}
\int d\bar{y}P[\tilde{x},\tilde{y};\varphi]\left\{ L[\tilde{y};\varphi]+\lambda N[\tilde{y};\varphi]+G(\tilde{y})\right\} =0\label{eq:8}
\end{equation}
see (\ref{eq:18-1}). Since then, the symbol $d\bar{y}$ will mean
that all the variables with tilda ($\tilde{}$) have the same time
$t$. 

From that, the arbitrary element $g$ in the expression (\ref{eq:6}):

\begin{equation}
g[\tilde{x};\varphi]=-\int d\bar{y}P[\tilde{x},\tilde{y};\varphi]\left\{ \lambda N[\tilde{y};\varphi]+G(\tilde{y})\right\} \label{eq:9}
\end{equation}
and the formula (\ref{eq:6}) can be described as this: 

\begin{equation}
L[\tilde{x};\varphi]-f[\tilde{x};\varphi]+\int d\bar{y}P[\tilde{x},\tilde{y};\varphi]\left\{ \lambda N[\tilde{y};\varphi]+G(\tilde{y})\right\} =0\label{eq:10}
\end{equation}
Here, $f[\tilde{x};\varphi]\in Q[\varphi]$$\tilde{F}_{\varphi}$.
Eq.\ref{eq:10} substituts Eq.\ref{eq:1} in the case of DCs (\ref{eq:3}),
which are satisfied by any solution to Eq.\ref{eq:10}. 

By comparison with Eq.\ref{eq:4}, the 'reaction force'

\begin{equation}
R[\tilde{x};\varphi]=f[\tilde{x};\varphi]+\int d\bar{y}Q[\tilde{x},\tilde{y};\varphi]\left\{ \lambda N[\tilde{y};\varphi]+G(\tilde{y})\right\} \label{eq:11}
\end{equation}

Following the analogy with classical mechanics one can say that Eq.\ref{eq:7}
resembles some surrogate of the \textbf{virtual work principle (VWP)}
- a surrogate because the 'reaction forces', at the moment t, can
not be perpendicular to the surface of the constraints (DC), Eq.\ref{eq:10}
resembles\textbf{ Lagrange's equations of the first kind}, and Eq.\ref{eq:11}
is a formula for the\textbf{ 'reaction forces'} of DC (\ref{eq:3}).
In this analogy, instead of the \textbf{virtual displacements}, we
have used appropriate linear projectors depending on the field $\varphi$.
The 'field' $\varphi$ in the simplest case may represent the radius
vector. But the main difference of presented approach to constraints
and canonical approach lies in the fact that there are explicitly
described rather acceleration restrictions caused by the presence
of constraints than constraint surfaces. See also\textbf{ \cite{Edwadia 2000}. }

\section{Classical constraints. Ideal constraints; A phenomenon of nonphysical
degrees of freedom}

We ask now how the restrictions (\ref{eq:3}) can be derived from
the classical contraints (CC) of the dynamical system (\ref{eq:1})?
To answer this question let us consider classical mechanics with the
constraints:

\begin{equation}
\sum a_{ij}(q,t)\dot{q}_{j}+g_{i}(q,t)=0\label{eq:12}
\end{equation}
where $\dot{q}_{j}$ is the j-th component of the vector $\dot{q}$.
Holonomic constraints can be differentiated once with respect to time
to get Eq.\ref{eq:12}. Differentiating once more with respect to
time, in both cases we get equations which, in the matrix-vector form,
are:

\begin{equation}
B(q,t)\ddot{q}=b(\dot{q},q,t)\label{eq:13}
\end{equation}
 The matrix $B$ in this equation has to be a singular. Otherwise,
it would be a dynamic equation, which for given initial conditions
would describe the problem in an unique way. If we assume that $B$
is a right invertible matrix, then such a right inverse exists that 

\begin{equation}
B(q,t)B_{R}^{-1}(q,t)=I\label{eq:14-1}
\end{equation}
 and 

\begin{equation}
B_{R}^{-1}(q,t)B(q,t)=Q(q,t)\label{eq:15-1}
\end{equation}
After multiplication of Eq.\ref{eq:13} by the inverse $B_{R}^{-1}(q,t)$
we get analoge of Eq.\ref{eq:3}. 

In fact, constraints equations obtained in the above way can have
the following structure:

\begin{equation}
Q'B(q,t)\ddot{q}=b(\dot{q},q,t)\label{eq:16-1}
\end{equation}
 with projected right invertible or invertible operator $B$, which
actually corresponds to a situation in which there are fewer constraints
than degrees of freedom, see App.3. Then, the equivalent equation:

\begin{equation}
Q\ddot{q}\equiv B_{R}^{-1}Q'B(q,t)\ddot{q}=B_{R}^{-1}b(\dot{q},q,t)\equiv f\label{eq:17-1}
\end{equation}
 has the form (\ref{eq:3}) with projector $Q=B_{R}^{-1}Q'B(q,t)$
, $L=\ddot{q}$ and the functional $f=B_{R}^{-1}b(\dot{q},q,t)$.
$Q$ indeed is a projector because: $B_{R}^{-1}Q'B(q,t)\cdot B_{R}^{-1}Q'B(q,t)=B_{R}^{-1}Q'B(q,t)\Longleftrightarrow Q^{2}=Q$.
See also App.3. 

Multiplying Eq.\ref{eq:3} with an operator depending on the field
$\varphi$:

\begin{equation}
A[\varphi]\Longleftrightarrow A[\tilde{x},\tilde{y};\varphi]\label{eq:18-1}
\end{equation}
we get equation:

\begin{equation}
A[\varphi]Q[\varphi]L[\varphi]=A[\varphi]f[\varphi]\label{eq:19-1}
\end{equation}
where $A[\varphi]$ and $Q[\varphi]$ operate in the space of functions
$\tilde{F}_{\varphi}\ni L[\varphi],f[\varphi]$. This equation is
equivalent to Eq.\ref{eq:3} if, for example, we assume that operator
$A[\varphi]$ is a right invertible:

\begin{equation}
A[\varphi]A[\varphi]_{R}^{-1}=I\label{20}
\end{equation}
 where $I$ is the unit operator in space $\tilde{F}_{\varphi}$,
and that

\begin{equation}
A_{R}^{-1}[\varphi]A[\varphi]=Q''\supseteq Q\label{eq:21-1}
\end{equation}
 where $Q'',Q$ are projectors.

\subsection{Ideal constraints}

In the case of ideal constraints in which the reaction forces $R_{ideal}[\tilde{x};\varphi]$
are perpendicular to the constraint surfaces and projectors $P_{ideal}[\tilde{x},\tilde{y};\varphi]$
projecting on the tangent surfaces at 'points' $\varphi(\tilde{x})$
are known, then we have, of course:

\begin{equation}
\int P_{ideal}[\tilde{x},\tilde{y};\varphi]R_{ideal}[\tilde{y};\varphi]d\bar{y}=0\label{eq:22-1}
\end{equation}
 In this case all derived formulas above will not be changed if

\begin{equation}
Q_{ideal}Q=Q_{ideal}\label{eq:23-1}
\end{equation}
 and $P_{ideal}=I-Q_{ideal}$but then, of course, the projector $Q$
has to be replaced by $Q_{ideal}$. 

\textbf{If (\ref{eq:23-1}) is not satisfied}, then, starting from
the formula (7), we have changes: so that (\ref{eq:8}) is modified
by 

\begin{equation}
\int d\bar{y}P_{ideal}[\tilde{x},\tilde{y};\varphi]\left\{ L[\tilde{y};\varphi]+\lambda N[\tilde{y};\varphi]+G(\tilde{y})\right\} =0\label{eq:24-1}
\end{equation}
 (\ref{eq:9}) is substituted by:

\begin{equation}
P_{ideal}g[\tilde{x};\varphi]=-P_{ideal}f[\tilde{x};\varphi]-\int d\bar{y}P_{ideal}[\tilde{x},\tilde{y};\varphi]\left\{ \lambda N[\tilde{y};\varphi]+G(\tilde{y})\right\} \label{eq:25-1}
\end{equation}
 and (\ref{eq:10}) is substituted by:

\begin{equation}
L[\tilde{x};\varphi]-Q_{ideal}f[\tilde{x};\varphi]+\int d\bar{y}P_{ideal}[\tilde{x},\tilde{y};\varphi]\left\{ \lambda N[\tilde{y};\varphi]+G(\tilde{y})\right\} =Q_{ideal}g[\tilde{x};\varphi]\label{eq:26-1}
\end{equation}
with arbitrary element $Q_{ideal}g[\tilde{x};\varphi]$. Now, the
'reaction forces' are:

\begin{eqnarray}
 & R[\tilde{x};\varphi]=\nonumber \\
 & Q_{ideal}f[\tilde{x};\varphi]+\int d\bar{y}Q_{ideal}[\tilde{x},\tilde{y};\varphi]\left\{ \lambda N[\tilde{y};\varphi]+G(\tilde{y})\right\} +Q_{ideal}g[\tilde{x};\varphi]\equiv\label{eq:27-1}\\
 & R_{ideal}[\tilde{x};\varphi]\nonumber \\
\nonumber 
\end{eqnarray}

The condition (\ref{eq:3}) leads to the restriction of the element
$QQ_{ideal}g[\tilde{x};\varphi]$:

\begin{equation}
QQ_{ideal}g[\tilde{x};\varphi]=f[\tilde{x};\varphi]-QQ_{ideal}f[\tilde{x};\varphi]+Q\int d\bar{y}P_{ideal}[\tilde{x},\tilde{y};\varphi]\left\{ \lambda N[\tilde{y};\varphi]+G(\tilde{y})\right\} \label{eq:28}
\end{equation}
 Substituting the formula (\ref{eq:28} in Eq.\ref{eq:26-1} the final
equation in which the ideal constraints (\ref{eq:22-1}) are incorporated
is the following:

\begin{eqnarray}
 & L[\tilde{x};\varphi]-f[\tilde{x};\varphi]-PQ_{ideal}f[\tilde{x};\varphi]+P\int d\bar{y}P_{ideal}[\tilde{x},\tilde{y};\varphi]\left\{ \lambda N[\tilde{y};\varphi]+G(\tilde{y})\right\} \nonumber \\
 & =PQ_{ideal}g[\tilde{x};\varphi]\label{eq:29}
\end{eqnarray}

\subsection{A phenomenon of nonphysical degrees of freedom}

Still, the component $PQ_{ideal}g[\tilde{x};\varphi]$ of the projection
$Q_{ideal}g[\tilde{x};\varphi]$ is unspecified. However, we claim
that Eq.\ref{eq:29} with arbitrary element $PQ_{ideal}g[\tilde{x};\varphi]=PQ_{ideal}Pg[\tilde{x};\varphi]$
correctly describes the problems with ideal constraints since in such
cases $PQ_{ideal}\asymp0$, see App.3. The above indeterminacy in
Eq.\ref{eq:29} occurs when we are outside of constraint surfaces.
This means that the system has fewer degrees of freedom than the number
of variables used in Eq.\ref{eq:29}. In another language we would
say that nonphysical degrees of freedom appear, which lead to the
presence of ambiguity of used formalism. Such ambiguity may affect
the results obtained, if some of the variables (nonphysical variables)
will not be expressed by the other, physical variables, according
to the constraints. This we call a \textit{phenomenon of nonphysical
degrees of freedom}. By the \textbf{\textit{physical variables}} we
understand here any minimal set of variables that are sufficient to
uniquely describe the configuration of the system in accordance with
the constraints (generalized variables).

\subsection{One general ideal constraint}

We illustrate the above process of thinking in the case of the one
general constraint (\ref{eq:54-1}) considered in App.3. In this case,
we can choose the following \textbf{symmetrical} projectors: 

\[
Q[\tilde{x},\tilde{y};\varphi]=\frac{R_{ideal}[\bar{x},t;\varphi]R_{ideal}[\bar{y},t;\varphi]}{\int R_{ideal}[\bar{z},t;\varphi]R_{ideal}[\bar{z},t;\varphi]d\bar{z}},\; P_{ideal}[\tilde{x},\tilde{y};\varphi]=\frac{\dot{\varphi}(\bar{x},t)\dot{\varphi}(\bar{y},t)}{\int\dot{\varphi}(\bar{z},t)\dot{\varphi}(\bar{z},t)d\bar{z}}
\]
 see (\ref{eq:60}) with Eq.\ref{eq:54}, and (\ref{eq:48-1}). Then,
on the constraint surface (\ref{eq:54-1}), we have:

\begin{equation}
QP_{ideal}=P_{ideal}Q\asymp0\label{eq:30}
\end{equation}
as pointed by using the symbol $'\asymp'$ instead of the strong equality
expressed by the symbol $'='$in the last equality. Strong equality
in the above formulas is the result of the symmetry of used projectors,
while weak equality results from the constraint equations: in this
case the vector $\dot{\varphi}$ is tangent and vector $R_{ideal}$
- normal to the constraint surface (\ref{eq:54-1}) - at the point
$\varphi$.

Using Eq.\ref{eq:5} and similar identity for projectors with subscript
'ideal', from (\ref{eq:30}), we have:

\begin{equation}
PQ_{ideal}=Q_{ideal}P\asymp0\label{eq:31}
\end{equation}
 and hence:

\begin{equation}
P\asymp PP_{ideal}\asymp P_{ideal}P\label{eq:32-1}
\end{equation}
 see App.3.

The first equalities of (\ref{eq:31}) and (\ref{eq:32-1}) simplify
equation (\ref{eq:29}) to the form:
\begin{equation}
L[\tilde{x};\varphi]-f[\tilde{x};\varphi]]+\int d\bar{y}P_{ideal}[\tilde{x},\tilde{y};\varphi]\left\{ \lambda N[\tilde{y};\varphi]+G(\tilde{y})\right\} =0\label{eq:33-1}
\end{equation}
 The resulting equation holds for any ideal constraints for which
Eqs (\ref{eq:30}) take place. This means that constraints (\ref{eq:3})
have to be described by specific, symmetric projectors $Q=I-P$ considered
in App.3.

\section{Examples of linear dynamical constraints (LDC)}

Let us collect the main results:

Eq.\ref{eq:10} is 

\[
L[\tilde{x};\varphi]-f[\tilde{x};\varphi]+\int d\bar{y}P[\tilde{x},\tilde{y};\varphi]\left\{ \lambda N[\tilde{y};\varphi]+G(\tilde{y})\right\} =0
\]

DC (\ref{eq:3}) are

\[
f[\tilde{x};\varphi]=QL[\tilde{x};\varphi]\equiv\int Q[\tilde{x},\tilde{y};\varphi]L[\tilde{y};\varphi]d\bar{y}
\]
with fixed functionals $L,f$. $P,Q$ - conjugate projectors (idempotent
operators) satisfying Eq.\ref{eq:5}: 

\[
P[\tilde{x},\tilde{y};\varphi]+Q[\tilde{x},\tilde{y};\varphi]=\delta(\tilde{x},\tilde{y})\Longleftrightarrow P+Q=I
\]

Because

\begin{equation}
PQ=QP=0,\; P=P^{2},\: Q=Q^{2}\label{eq:18}
\end{equation}
 we see that DC (\ref{eq:3}) result immediately from Eq.\ref{eq:10}.

Let us take DC (\ref{eq:3}) with 

\begin{equation}
f[\tilde{x};\varphi]=\mu QL[\tilde{x};\varphi]\label{eq:19}
\end{equation}
 Hence and from Eq.\ref{eq:10} 

\begin{equation}
(I-\mu Q)L[\tilde{x};\varphi]+\int d\tilde{y}P[\tilde{x},\tilde{y};\varphi]\left\{ \lambda N[\tilde{y};\varphi]+G(\tilde{y})\right\} =0\label{eq:20}
\end{equation}
 We can tell immediately that, at $\mu=1$, DC (\ref{eq:19-1}), (\ref{eq:3})
lead only to weakning of the original Eq.\ref{eq:1}.What happens,
for $\mu\neq1$? In this case, by inverting the operator $I-\mu Q$,
we get the following equation:

\begin{eqnarray}
 & L[\tilde{x};\varphi]+(I-\mu Q)^{-1}\int d\bar{y}P[\tilde{x},\tilde{y};\varphi]\left\{ \lambda N[\tilde{y};\varphi]+G(\tilde{y})\right\} =\nonumber \\
 & L[\tilde{x};\varphi]+\int d\bar{y}P[\tilde{x},\tilde{y};\varphi]\left\{ \lambda N[\tilde{y};\varphi]+G(\tilde{y})\right\} =0\label{eq:21}
\end{eqnarray}
 One can understand this result if we take into account that now$\int Q[\tilde{x},\tilde{y};\varphi]L[\tilde{y};\varphi]d\tilde{y}=0$. 

Another example of the linear DC (\ref{eq:3}) is given by:

\begin{equation}
f[\tilde{x};\varphi]=-QM\varphi(\tilde{x})\equiv-\int d\bar{y}d\tilde{z}Q(\tilde{x},\tilde{y})M(\tilde{y},\tilde{z})\varphi(\tilde{z})\label{eq:22}
\end{equation}
 where $M$ is a given constant matrix and $Q$ a projector, both
independent of $\varphi$. Now, Eq.\ref{eq:10} is given by

\begin{equation}
L_{M}[\tilde{x};\varphi]+\int d\tilde{y}P(\tilde{x},\tilde{y})\left\{ \lambda N[\tilde{y};\varphi]+G(\tilde{y})\right\} =0\label{eq:23}
\end{equation}
with the linear functional: $L_{M}[\tilde{x};\varphi]=L[\tilde{x};\varphi]+QM\varphi(\tilde{x})$.
The term $QM\varphi$ can describe parameters which do not appear
in the first term because, e.g., of symmetry in a certain area of
considered equations. 

One can finally say that \textbf{any knowledge about the main linear
term of Eq.\ref{eq:1},} expressed in the form of (\ref{eq:3}), allows
us to change this equation to the form of Eq.\ref{eq:10} if the analoge
of the virtual work principle is assumed. In this way, relying more
on observation than on the proliferation of some ideas, you can try
to understand some phenomena.

\section{Examples of nonlinear dynamical constraints (NDC) and linear original
dynamics (LOD)}

We assume that, for example:

\begin{equation}
f[\tilde{x};\varphi]=\mu QN[\tilde{x};\varphi]\Longleftrightarrow f[\tilde{x};\varphi]=\mu\int Q[\tilde{x},\tilde{y};\varphi]N[\tilde{y};\varphi]d\bar{y}\label{eq:24}
\end{equation}
 This effectively means that, for $G=0$, we modify the nonlinear
part of Eq.\ref{eq:1}. In this case, Eq.\ref{eq:10} takes the form:

\begin{equation}
L[\tilde{x};\varphi]-\mu\int Q[\tilde{x},\tilde{y};\varphi]N[\tilde{y};\varphi]d\bar{y}+\int d\bar{y}P[\tilde{x},\tilde{y};\varphi]\left\{ \lambda N[\tilde{y};\varphi]+G(\tilde{y})\right\} =0\label{eq:25}
\end{equation}
 or in short as 

\begin{equation}
L+\left(\lambda P-\mu Q\right)N+PG=0\label{eq:26}
\end{equation}
 Hence, the equivalent, 

\begin{equation}
\left(\lambda P-\mu Q\right)^{-1}(L+PG)+N=0\label{eq:27}
\end{equation}
 where $\left(\lambda P-\mu Q\right)^{-1}=\lambda^{-1}P-\mu^{-1}Q$,
see (\ref{eq:18-1}). In other words, all the modification of the
theory can be transferred to linear terms, although without non-linear
terms the above modification disappears!

In all these examples one can treat some constant of motions as constatraints
related to a kind of material surfaces and some, as energy, as purely
dynamical quantities, see also App.1.

\subsection{Linear original dynamics}

In this case, the starting equation is:

\begin{equation}
L[\tilde{x};\varphi]+G(\tilde{x})=0\label{eq:40-2}
\end{equation}
 It is an inhomogeneous linear equation which should be changed to
satify the constraints (\ref{eq:3}). In fact, we only have to modify
previously derived formulas putting the nonlinear term $N\equiv0$.
It results that, for the ideal constraints, the modified formulas
are nonlinear even for the linear constraints, see (\ref{eq:48-1})
and Eq.\ref{eq:29}.

\section{Appendix}

\subsection{About analogy with classical mechanics and the essence of the constraints}

In classical mechanics the main quantity around which everything revolves
is an accerelation of objects either extended or point like particles.
The accelerations in the dynamical equations appear in the linear
way. Moreover, if the constrains are proper times differentiated with
respect to time (once or twice), then accelerations also appear in
the linear way, see \cite{Edwadia 1993}. Such quantities, which describe
changes, or changes of changes as in the case of acceleration, also
appear in a linear way in the case of 'physical' fields describing
extended systems. They are responsible for the additional conditions
as the initial and boundary conditions, which must be taken into account
to get an unique solution to the considered equations. 

When we look at constraints as constants of motion, the question naturally
arises, what is the difference? The difference lies in the fact that
other constants of motion are not carried out by the physical surfaces
as the constants of motion interpreted as the constraints. 

Constants of motion related to constraints are explicitly present
in the theory. Constrains, however, limit the initial conditions of
theory and failure to do so push the system outside the surface of
the constraints. They also are related to a global description of
systems. Constrains have an effect on local interaction of individuals
composing complex systems like economic systems.

\subsection{About the classical and dynamical constraints ((CC) and(DC))}

By CC we understand mathematical or physical restrictions which descriptions
does not require 'accelerations' or their analogues. In classical
mechanics they are called holonomic and nonholonomic constraints.
From their definitions results that for systems with CC the initial
and boundary conditions can not be arbitrary. It is result of fact
that CC eliminate some number of degrees of freedom like in the case
of pendulum or incompressible liquid. 

Main difference with DC is such that CC are automatically realized
by there equations: 'surfaces' which realizes such constraints. \textbf{This
is not the case of DC which are realized by the extra forces calculated
with the help of dynamical equations! }

\subsection{Spherical and more general ideal constraints}

Let us assume that we have the following spherical constraint: 

\begin{equation}
\int d\overline{y}\varphi^{2}(\bar{y},t)=constant=R^{2}\label{eq:32}
\end{equation}
 Hence,

\begin{equation}
\int d\bar{y}\varphi(\bar{y},t)\dot{\varphi}(\bar{y},t)=0\label{eq:33}
\end{equation}
and 

\begin{equation}
\int d\bar{y}\varphi(\bar{y},t)\ddot{\varphi}(\bar{y},t)+\int d\bar{y}\dot{\varphi}(\bar{y},t)\dot{\varphi}(\bar{y},t)=0\label{eq:34-1}
\end{equation}

In this case, to get an analoge of formula (\ref{eq:13}), or rather
(\ref{eq:16-1}), we can choose: 

\begin{equation}
B[\bar{x},\bar{y},t;\varphi]=\delta(\bar{x}-\bar{y})\varphi(\bar{x},t)\label{eq:34}
\end{equation}
 where $B$ is a nonsingular operator at least for t for which $\varphi\neq0$:

\begin{equation}
B_{R}^{-1}[\bar{y},\bar{z},t;\varphi]=B^{-1}[\bar{y},\bar{z},t;\varphi]=\delta(\bar{y}-\bar{z})\frac{1}{\varphi(\bar{y},t)}\label{eq:35}
\end{equation}

\begin{equation}
Q'(\bar{x},\bar{y})=\frac{1}{V}\int d\bar{x}\delta(\bar{x}-\bar{y})=\frac{1}{V}\label{eq:36}
\end{equation}
and 

\begin{equation}
b[t;\varphi]=-\int d\bar{y}\dot{\varphi}(\bar{y},t)\dot{\varphi}(\bar{y},t)\label{eq:38-1}
\end{equation}
Here V denotes the volume of an integration region, $\bar{x}\in V$.
Of course, (\ref{eq:36}) is a projector, which action on a function
is reduced to integration and multiplication by the factor $1/V$
to get in result a constant. Now, we can use the formula (\ref{eq:16-1})
and (\ref{eq:17-1}) to describe CC (\ref{eq:32-1}) in the form of
Eq.\ref{eq:3} of DC with

\begin{eqnarray}
 & Q[\bar{x},\bar{y},t;\varphi]=B_{R}^{-1}Q'B[\bar{x},\bar{y},t;\varphi]=\int B_{R}^{-1}[\bar{x},\bar{z},t;\varphi]Q'(\bar{z},\bar{w})B[\bar{w},\bar{y,t};\varphi]=\nonumber \\
 & \int d\bar{z}d\bar{w}\delta(\bar{x}-\bar{z})\frac{1}{\varphi(\bar{x},t)}\frac{1}{V}\int d\bar{z}'\delta(\bar{z}'-\bar{w})\delta(\bar{w}-\bar{y})\varphi(\bar{w},t)=\frac{1}{V}\frac{\varphi(\bar{y},t)}{\varphi(\bar{x},t)}\label{eq:37}
\end{eqnarray}

Hence, in the DC (\ref{eq:3}), 

\begin{equation}
f[\bar{x},t;\varphi]=-\frac{1}{\varphi(\bar{x},t)}\int d\bar{y}\dot{\varphi}(\bar{y},t)\dot{\varphi}(\bar{y},t)\label{eq:40-1}
\end{equation}

It is worth noting here that Q is a projector, but it is a symmetric
projector only for all field variables equal to each other:

\begin{equation}
\varphi(\bar{x},t)=\varphi(\bar{y},t),\; for\:\bar{x},\bar{y}\in V\label{eq:38}
\end{equation}
In other cases, (\ref{eq:3}) and (\ref{eq:7}), with (\ref{eq:37}),
can describe the \textbf{non-ideal constraints} described by a \textbf{surrogate
of virtual work principle}:

\begin{equation}
PR[\bar{x},t;\varphi]=R[\bar{x},t;\varphi]-\frac{1}{V\varphi(\bar{x},t)}\int d\bar{y}\varphi(\bar{y},t)R[\bar{y},t;\varphi]=0\label{eq:39}
\end{equation}
 where the projector $P$ was chosen as:

\begin{equation}
P[\bar{x},\bar{y},t;\varphi]=\delta(\bar{x}-\bar{y})-Q[\bar{x},\bar{y},t;\varphi]=\delta(\bar{x}-\bar{y})-\frac{1}{V}\frac{\varphi(\bar{y},t)}{\varphi(\bar{x},t)}\label{eq:40}
\end{equation}
 This projector reflects circular symmetry in the case of non-ideal
costraints (\ref{eq:32-1}). 

From (\ref{eq:39}), 

\begin{equation}
R[\bar{y},t;\varphi]=\frac{G[t;\varphi]}{\varphi(\bar{y},t)}\label{eq:41}
\end{equation}
 where a functional $G$ does not depend on the variable $\bar{y}.$
The values of 'field' in the denominator should not necessarily worry
us, because the infinity of the expression $1/\varphi(\bar{y},t)$
, for $t\rightarrow t'$, can be simultaneosly neutralized by $G\rightarrow0$
. 

Once more, for the ideal spherical constraints, where a sphere is
considered in the space $F$ of fuctions $\varphi$, we should have:

\begin{equation}
R[\bar{y},t;\varphi]=H[t;\varphi]\varphi(\bar{y},t)\label{eq:47-1}
\end{equation}
 with a fuctional $H[t;\varphi]$ which do not depend on variable$\bar{y}$.
From (\ref{eq:41}) we get

\[
\varphi(\bar{y},t)^{2}=\frac{G[\varphi]}{H[\varphi]}
\]
 but this would mean that $\varphi$ does not depend on $\bar{y}$
in a continuous way. It also means that in this case the conditions
(\ref{eq:39}) and (\ref{eq:3}) can not describe ideal constraints. 

Spherical constraints describe the simplest nonlinear, holonomic constraints
in physics. They contain the symmetry of the circle, which throughout
human history has been synonymous with - excellence. So would not
be strange if they would be found in some basic field theory describing
the Universe. A sphere in the configuration space of such system as
the universe is the favorite model in cosmology. It was also considered
by Henri Poincare, see Wikipedia. In fact, the constraints (\ref{eq:32-1})
do not mean that all particles are located on the sphere with radius
$R$ but only that the sum of all squars of their radius vectors is
equal to $R^{2}$. They can describe a fany model of particles in
which location of one particle at the extreme distance equal to R
leads to locations of other particles at the center of the sphere
with radius $R$! In other words, in this model the influence of the
global quantity represented by Eq.\ref{eq:32-1} on the local inter-particle
interaction can be traced.

\subsubsection{A single sclerenomic ideal constraint}

\begin{equation}
H[\varphi]=constant\label{eq:54-1}
\end{equation}
In this case the reaction forces are proportional to the gradient
of the functional $H$: 

\begin{equation}
R_{ideal}[\bar{y};\varphi]\propto\frac{\delta H[\varphi]}{\delta\varphi(\bar{y},)}\equiv V[\bar{y};\varphi]\equiv V\label{eq:54}
\end{equation}
describes a given constraint surface. $H$ may depend on the 'space'
variable $\bar{y}$ of the fuction $\varphi$ in the non-local way,
see Eq.\ref{eq:32-1}. Like in classical mechanics we assume that
'all' $\varphi$ are taken at the same time $t$. We also assume that
the functional derivative $\delta/\delta\varphi(\bar{y})$ is defined
in such a way that $\delta\varphi(\bar{x})/\delta\varphi(\bar{y})=\delta(\bar{x}-\bar{y})$.
Let us also notice that from (\ref{eq:54-1})

\begin{equation}
\frac{d}{dt}H[\varphi(t)]=\int d\bar{y}V[\bar{y};\varphi(t)]\dot{\varphi}(\bar{y},t)=0\label{eq:61}
\end{equation}
 The above equation resulting from the observation that $H$ is also
a constant of motion, see (\ref{eq:54-1}), shows to us that $R_{ideal}$defined
by Eq.\ref{eq:54} is perpendicular to the surface $H$ at the 'point'
$\varphi$. 

For the projector

\begin{equation}
P_{ideal}=\frac{\dot{\varphi}(\bar{x},t)\dot{\varphi}(\bar{y},t)}{\int\dot{\varphi}(\bar{z},t)\dot{\varphi}(\bar{z},t)d\bar{z}}\label{eq:48-1}
\end{equation}
 where $\dot{\varphi}(\bar{x},t)=\frac{\partial}{\partial t}\varphi(\bar{x},t)$,
we have

\begin{equation}
\int P_{ideal}[\tilde{x},\tilde{y};\varphi]R_{ideal}[\tilde{y};\varphi]d\bar{y}=0\label{eq:55}
\end{equation}
Hence, we can interprete the projector $P_{ideal}\equiv P_{ideal}[\bar{x},\bar{y},t;\varphi]$
as an operator projecting on the tangent space of the surface \ref{eq:54-1}
at the 'point' $\varphi$. 

We have to remind you that in the all above formulas, the symbol $d\bar{y}$
means that in vectors $\tilde{x},\tilde{y},\tilde{z}$ all time components
are equal to $t$. In the case of general, scleronomic (explicitly
independent of time) ideal constraints described by the Eq.\ref{eq:54-1},
by double differentations, we get an equation similar to Eq.\ref{eq:13}:

\begin{equation}
\int d\bar{y}V[\bar{y};\varphi(t)]\ddot{\varphi}(\bar{y},t)=b[\dot{\varphi},\varphi,t]\label{eq:58}
\end{equation}
 which can be described in an equivalent form as follows:

\begin{equation}
V[\bar{x};\varphi(t)]\int d\bar{y}V[\bar{y};\varphi(t)]\ddot{\varphi}(\bar{y},t)=V[\bar{x};\varphi(t)]b[\dot{\varphi},\varphi,t]\equiv f[\tilde{x};\varphi]\label{eq:59}
\end{equation}
This is a rather peculiar equivalent form of Eq.\ref{eq:58}, but
thanks to the above substitution the constraints (\ref{eq:54-1})
can be described in the form of Eq.\ref{eq:3}.This can be seen if
the symmetric projector 

\begin{equation}
Q[\tilde{x},\tilde{y};\varphi]=\frac{V[\bar{x};\varphi(t)]V[\bar{y};\varphi(t)]}{\int V[\bar{z};\varphi(t)]V[\bar{z};\varphi(t)]}\label{eq:60}
\end{equation}
 is introduced $\spadesuit$. For definition of $V[\bar{y};\varphi]$,
see Eq.\ref{eq:54}. Further generalization of this topic see just
below.

\subsubsection{A few sclerenomic ideal constraints}

In such cases, instead of a single Eq.\ref{eq:54-1} we have more
equations:

\begin{equation}
H_{i}[\varphi]=constant,\quad for\; i=1,2...,k\label{eq:67}
\end{equation}
 To each of them one can write:

\begin{equation}
\frac{d}{dt}H_{i}[\varphi(t)]=\int d\bar{y}\frac{\delta H_{i}[\varphi(t)]}{\delta\varphi(\bar{y},t)}\dot{\varphi}(\bar{y},t)\equiv\int d\bar{y}V_{i}[\bar{y};\varphi]\dot{\varphi}(\bar{y})\asymp0\label{eq:68}
\end{equation}
 where the symbol $'\asymp'$ means that Eqs (\ref{eq:68}) are satisfied
only if $\varphi$ fulfils constrain equations (\ref{eq:67}). 

Acting again with the time derivative on the (\ref{eq:68}), we get

\begin{equation}
\int d\bar{y}V_{i}[\bar{y};\varphi]\ddot{\varphi}(\bar{y})\asymp f_{i}[\varphi,\dot{\varphi}]\label{eq:69-1}
\end{equation}
for i=1,...,k. Goal that we set now is: How constraints (\ref{eq:67})
descibed in the form (\ref{eq:69-1}) can be written in the form of
the Eq.\ref{eq:3} with symmetrical projector $Q$? For the sake of
simplicity, we assume that

\begin{equation}
L[\tilde{x}:\varphi]\equiv\ddot{\varphi}(\tilde{x})=\ddot{\varphi}(t,\bar{x},\alpha,\beta,\gamma,...)\label{eq:70-1}
\end{equation}

Let us treat these $k$ f-f $V_{i}[\bar{y};\varphi]$ as k vectors
denoted by $V_{i}$:

\begin{equation}
V_{i}\Longleftrightarrow V_{i}[\bar{y};\varphi]=\frac{\delta H_{i}[\varphi]}{\delta\varphi(\bar{y})}\label{eq:69}
\end{equation}
 Then, (\ref{eq:68}) is:

\begin{equation}
<V_{i},\dot{\varphi}>\asymp0\label{eq:70}
\end{equation}
 for i=1,...,k. These equations tell us that at the 'point' $\varphi$
the vectors $V_{i}$ are perpendicular to the vectors $\dot{\varphi}$,
the infinitesimal change of which to the infinitesimal change of time
is tangent to the constraints surface. If vectors $V_{i}$, which
also enter Eq.\ref{eq:69-1}), are linear independent, then by means
of Gram-Schmidt process one can construct $k$ orthonormal vectors
$U_{i}$ which are also perpendicular to the velocity vectors $\dot{\varphi}$.
With the help of them and Dirac's notation one can express the projector
$Q$ of Eq.\ref{eq:3} as follows:

\begin{equation}
Q=\sum_{i=1}^{k}|U_{i}><U_{i}|\Longleftrightarrow Q[\bar{x},\bar{y};\varphi]=\sum_{i=1}^{k}U_{i}[\bar{x};\varphi]U_{i}[\bar{y};\varphi]\label{eq:71}
\end{equation}
 where we have assumed that all values of functions $U_{i}$are real.
This is a generalization of formula (\ref{eq:60}). The projector
$P=I-Q$ projects on space of vectors tangent to the constraint surface
at the point $\varphi$. One can also show that the above projector
$Q$ constructed by means of the Gram-Schmidt process satisfies 

\begin{equation}
QP_{ideal}=P_{ideal}Q\asymp0\label{eq:72}
\end{equation}
Hence, introducing the pair of projectors $P_{ideal}+Q_{ideal}=I$
, where $P_{ideal}$is given by the formula (\ref{eq:48-1} ), one
can derive the following equalities:

\begin{equation}
Q\asymp Q_{ideal}QQ_{ideal}\; and\; P_{ideal}\asymp PP_{ideal}P\label{eq:73}
\end{equation}
which allows to make the following identification of projectors:

\begin{equation}
Q\asymp Q_{ideal},\quad P\asymp P_{ideal}\label{eq:74}
\end{equation}
 The above identifications allow us to satisfy Eqs (\ref{eq:73})
as well Eqs (\ref{eq:72}). However, to describe Eq.\ref{eq:29} as
an equation (\ref{eq:33-1}) the scleronomic constraints (\ref{eq:67})
should be described in the form (\ref{eq:3}) with the help of the
projector (\ref{eq:74}) constructed with Gram-Schmidt process. See
Sec.3 (A phenomenon of nonphysical degrees of freedom)

To see that conditions (\ref{eq:70}) are also satisfied by orthogonal
vectors $U_{i}'$ obtained from vectors $V_{j}$ in the Gram-Schmidt
process I will write them here with the help of function:

\[
proj_{U}(V)=\frac{<U,V>}{<U,U>}U
\]
 and the reccurent formula

\[
U_{1}'=V_{1},
\]

\[
U_{j}'[\bar{x};\varphi]\Longleftrightarrow U_{j}'=V_{j}+\sum_{i=1}^{j-1}proj_{U_{i}'}(V_{j})
\]
for j=2,...,k. It is easy to see that the projector $Q$ constructed
by means of normalized vectors, $U_{j}=U_{j}'/<U_{j}',U_{j}'>$, via
the formula (\ref{eq:71}), satisfies conditions (\ref{eq:72}).

\subsection{About one-sided constraints (CC) in classical mechanics; short-range
forces}

On this subject I speak of the following reasons: First, in the Internet,
I found the discussion of such constraints by means of advanced means
or complicated cases including solid mechanics. Secondly, as previously
discussed, I am focusing not on the elimination of redundant degrees
of freedom, but on the forces that are doing it. 

In the case of n material points, the one-sided constraints are characterized
not by equations but by inequalities. Thus, in the case of holonomic
constraints we hawe:

\[
f_{i}(\vec{r}_{1},...,\vec{r}_{n};t)\leq0,\; for\: i=1,...,k<3n
\]
where $\vec{r}_{i}$ means the radius vector of the i-th particle.
Inequalities mean a drastic loosening of restrictions: only if there
is 'threat' of their failure, the system 'suffers' of reaction forces.
Such situation can be described by short-range forces, whose centers
satisfy the equations

\[
f_{i}(\vec{r}_{1},...,\vec{r}_{n};t)=0,\; for\: i=1,...,k<3n
\]
 Usually, the surfaces satisfying the above equations are calle the
\textit{walls}. Short-range reaction forces should be a priori chosen
in such a way that an energy, which is available for individual particles
is not enough to cross the walls. In this way we avoid tracking, when
the particles are hiting in to the walls, nor the need for discontinuous
changes in their momenta. Everything is encoded in the dynamical equations.

\subsection{About one-sided invertible operators }

A\textit{ right invertible operator} $A$ is defined as an operator
for which one can write the following equation:

\begin{equation}
AA_{R}^{-1}=I\label{eq:46}
\end{equation}
 with not uniquely chosen a right inverse operator $A_{R}^{-1}$ and
the unite operator $I$ in a considered linear space. For a left invertible
operator, we would have a similar definition, but the operator $A_{R}^{-1}$is
substituted by an operator $A_{l}^{-1}$ standing at the l.h.s. of
the operator $A$:

\begin{equation}
A_{l}^{-1}A=I\label{eq:47}
\end{equation}
Occurring here operators $A_{R}^{-1},A_{l}^{-1}$satisfy the first
two demands of the Moore-Penrose definition of the generalized inverse
(pseudoinverse) denoted by $A^{+}$:

\begin{equation}
(1)\; AA^{+}A=A\label{eq:48}
\end{equation}

\begin{equation}
(2)\; A^{+}AA^{+}=A^{+}\label{eq:49}
\end{equation}
 and often, in considered examples, are satisfied the second two demands
of the Moore-Penrose definition,:

\begin{equation}
(3)\;\left(AA^{+}\right)^{*}=AA^{+}\label{eq:50}
\end{equation}
 
\begin{equation}
(4)\;\left(A^{+}A\right)^{*}=A^{+}A\label{eq:51}
\end{equation}
 see \cite{Hanc 2012}, what guarantees of getting a least squares
solution to the considered system of equations. 

We think however that one-sided invertible operators in the sense
of Eqs(\ref{eq:46},\ref{eq:47}), are more simple and therefore more
useful for basic description of nature, and except that, the request:
'least squares solution' is not always necessary. see \cite{Przew. 1988,Hanc 2012}
and other author's 'recent' papers.

\subsection{About strange behavior of some objects}

Let us assume that we consider a discrete system such that from Eq.\ref{eq:4}
$R$, the reaction force, has to be a vector. In this case Eq.\ref{eq:41}
means that $G$ is not a scalar but must behave so that $R$ , at
the transformation of the coordinate system, is the vector. Taking,
however, the scalar product of the two vectors $\varphi,R$ :

\begin{equation}
\left(\varphi(\cdot,t),R[\cdot,t;\varphi]\right)=VG[t;\varphi]\label{eq:52}
\end{equation}
 we should get, in the r.h.s., the scalar. This explicit contradiction,
we can probably explained by the fact that $G$ behaves as a scalar
on the subset of vectors $\varphi$ satisfying Eq.\ref{eq:32-1}.

\subsection{About space $\tilde{M}$, Cantor's theorem and evolution theory}

In Sec.1 we said that the set $\tilde{M}$ consists of elements (vectors)
reflecting specific properties of the considered system. This is only
partly true because in these elements are also included certain properties
of the observer as the experience of one, two or three dimensional
spaces. As we know from the Cantor's theorem, there is 1-1 correspondence
between the points of the plane or of n-dimensional space and of the
stright line. It seems, however, that the identification of objects
with a higher dimensional space is \textbf{much simpler and effective}
than using the one-dimensional, and this was used at least by some
organisms, see also \cite{Nerlich 1994}, page 20, where other opinions
are presented. 

Higher dimensional spaces particularly preferred by quantum field
theory to get meaningful theory appear to be evidence of the fact
that even in the field of logic a similar phenomenon can be observed.
By means of constraints certain dimensions can be roll up. By means
of them also some constants can be introduced into considered equations.

\end{document}